\documentclass[twocolumn,showpacs,preprintnumbers,amsmath,amssymb]{revtex4}
\usepackage{dcolumn}
\usepackage{bm}
\usepackage{graphicx}
\usepackage{amsmath}
\begin{document}

\title{Regular and chaotic Bose-Einstein condensate in an accelerated Wannier-Stark lattice}
\author{Wenhua Hai, Gengbiao Lu,  Honghua Zhong}
\affiliation{Key Laboratory of Low-dimensional Quantum Structures
and Quantum Control of Ministry of Education, and
\\ Department of Physics, Hunan Normal University, Changsha 410081, China}

\begin{abstract}

We investigate a Bose-Einstein condensate held in a
quasi-one-dimensional Wannier-Stark lattice which is a combination
of linear potential with an accelerated optical lattice. It is
demonstrated that the system can be reduced to a periodically driven
Gross-Pitaevskii one, in which we find the first exact analytical
solution and the regular and chaotic numerical solutions with
accelerated atomic flow densities. The results suggest an
experimental scheme for generating and controlling the accelerating
regular and chaotic matter-waves.

\end{abstract}


\maketitle

\section{Introduction}

Recently, experimental investigation on the Bose-Einstein
condensates (BECs) in a Wannier-Stark (WS) system have attracted
much attention \cite{Anderson}. In some experiments, a BEC of
rubidium atoms was created, which then was loaded into the optical
lattice potential of a standing laser light \cite{omorsch} and was
adopted to observe the collective tunneling effects \cite{Anderson}.
The optical potentials have been used in many theoretical and
experimental studies of quantum dynamics, for example, the coherent
pulse output from BECs in WS system \cite{MGl}, the observation of
Bloch oscillations both with single atoms and with a BEC in an
accelerated standing wave \cite{omorsch, mben}, the studies of WS
ladders for the accelerated optical potential \cite{SRW}. An
atomic-scale analog of the kicked rotor is realized by placing
laser-cooled atoms in a pulsed standing wave \cite{QTho}. Very
recently, M. Gl\"{u}ck et al. investigated the properties of a
coherent superposition of WS resonances \cite{MGl} and gave the
lifetime of WS states \cite{Mglu}. Chaotic behaviors have also been
found in the WS systems without acceleration \cite{QTho, Thommen,
Mglu2, Fang}. It is a natural motivation for us to demonstrate the
chaotic and regular features for the BEC in an accelerated WS
lattice.

On the other hand, it is well-known that the BEC governed by a
Gross-Pitaevskii equation (GPE) without external potential is an
integrable system and the integrability could be easily broken by
external potentials of different forms \cite{Zhao}. So previously,
only few analytical works concern exact solutions of the system,
where one-dimensional (1D) stationary systems with some simple
potentials are treated, such as the infinite or finite square-wells
\cite{Agosta,Carr,Leboeuf,Kagan}, the step-potentials \cite{Seaman},
$\delta$ or $\delta$ comb potentials \cite{Witthaut,Hakim,BTSeaman},
linear ramp potential \cite{Khawaja, Tuszynski} and optical lattice
potentials \cite{Bronski, Bronski2}. Under some rigorous conditions
on the interaction intensities or external potentials, several exact
nonstationary-state solutions were constructed \cite{Deconinck,
Hai}, including the exact soliton solutions \cite{Beitia, Beitia2,
Liang, Chong, Hai1}. It is worth noting that the balance between
nonlinearity and dispersion was found in the seminal work of soliton
\cite{Zabusky}, and the new balances between the atom-atom
interaction and the external potentials are demonstrated recently
\cite{Rodas, Hai, Hai2}. By using the balance conditions, although
some exact solutions have been constructed for the GPE with periodic
potential, however, any exact solution in the nonintegrable WS
system with a combination of the linear and periodic potentials has
not been reported yet.

The aim of this paper is to present the first exact analytical
solution with balance condition and to illustrate the regular and
chaotic numerical solutions of the accelerated WS system. The
corresponding atomic flow densities accelerated by the constant
force are demonstrated. Based on the relations between the system
parameters and the solution behaviors, we suggest an experimental
method for controlling the regular and chaotic states by applying
the accelerated optical potential and adjusting the system
parameters.

\section{Simplification of the Wannier-Stark system}

The mean-field theory is a successful one for describing the BECs.
In this theory, the dynamical behaviors are governed by the GPE
\cite{Dalfovo, Leggett}, which provides us a nonlinear macroscopic
quantum system. Let us consider a BEC trapped in one-dimensional
tilted optical lattice potential
\begin{eqnarray}
V(x',t')=V_0\cos(2k_L\xi')+F x',\ \ \ \ \xi'=x'+\frac 1 2 a t'^2;
\end{eqnarray}
here $x'$ and $t'$ are spatial and time coordinates,
$V_0\cos(2k_L\xi')$ is the accelerating optical potential \cite{SRW,
Thommen} with strength $V_0$, wave vector $k_L$ and acceleration
$a$, and $F=m a$ is a constant force \cite{Thommen} with $m$ being
the atomic mass. Due to this force, a ``tilted" potential is
produced that leads the atoms to accelerate in $x$ direction with
linearly increasing flow density and makes the atoms tunnel out of
the optical traps. The corresponding dimensionless GPE reads as
\begin{eqnarray}
i\frac{\partial \psi}{\partial t}=- \frac{\partial^2\psi}{\partial
x^2}+[V_0\cos (2\xi)+\alpha x+g_{1d}|\psi|^{2}]\psi,
\\ \xi=k_L\xi'=x+\alpha t^2,\ \ \ \alpha=\frac 1 2 k_L a
\hbar^2/E_r^2, \nonumber
\end{eqnarray}
where the dimensionless spatial and time coordinates are $x=k_L x'$
and $t=E_r t'/\hbar$. The wave function $\psi$ has been normalized
in units of the radical $\sqrt {k_L}$, the potential depth $V_{0}$
is normalized by the recoil energy $E_r=\hbar^2k_L^2/(2m)$. The
constant force is rescaled by the unit $k_LE_r$, so all variables
and parameters in equation (3) are dimensionless. In such units, the
interatomic interaction intensity related to the $s$-wave scattering
length $a_{s}$ is in the form $g_{1d}=4 a_{s}/(k_L l_r^2)$ with
$l_r=\sqrt{\hbar/(m\omega_r)}$ being the radial length of harmonic
oscillator.

In order to get a simple description and better understanding of BEC
dynamics, we let the wave function be in the form
\begin{eqnarray}
\psi(x,t)=u(\xi, t)\exp\Big[-i\Big(\alpha x t+\frac 1 3
\alpha^2t^3\Big)\Big],
\end{eqnarray}
where undetermined function $u(\xi, t)$ may be real or complex,
which is normalized to the total number of atoms, $\int
u^{2}(\xi,t)dx=N$. Given Eq. (2), we perform the calculations
\begin{eqnarray}
i\frac{\partial \psi}{\partial t}&=& i\Big(\frac{\partial
u}{\partial t}+2\alpha t\frac{\partial u}{\partial
\xi}\Big)e^{-i(\alpha x t+\frac 1 3 \alpha^2t^3)}
+(\alpha x+ \alpha^2t^2)\psi, \nonumber \\
\frac{\partial^2 \psi}{\partial x^2}&=& \Big(\frac{\partial^2
u}{\partial \xi^2}-i2\alpha t\frac{\partial u}{\partial
\xi}\Big)e^{-i(\alpha x t+\frac 1 3 \alpha^2t^3)}-\alpha^2t^2\psi.
\end{eqnarray}
Substituting Eq. (4) into Eq. (2) yields
\begin{eqnarray}
i\frac{\partial u}{\partial t}=- \frac{\partial^2u}{\partial
\xi^2}+[V_0\cos (2\xi)+g_{1d}|u|^{2}]u
\end{eqnarray}
in which the linear potential is removed and its effect is included
in the parameter $\alpha$. When the optical potential is switched
off, $V_0=0$, Eq. (5) becomes a standard nonlinear Schr\"{o}dinger
equation (NLSE), whose single-soliton and multisoliton solutions are
well-known for us. It is worth noting that in the transformation
$x\to x+n\pi$ for $n=0,1,2,\cdots$, Eq. (5) and its solution
$u(\xi,t)$ are kept such that Eq. (3) gives $n$ solutions
\begin{eqnarray}
\psi_n(x,t)=\psi(x+n\pi,t)=\psi(x,t)e^{-i\alpha n\pi t}.
\end{eqnarray}
The solutions with different $n$ possess the different phases and
the same amplitude.

\section{Exact periodic wave with accelerated flow}

We are interested in the exact analytical solution and regular and
chaotic numerical solutions of Eq. (5). Noticing that $\xi$ and $t$
in Eq. (5) are two independent variables, we can rewrite the
function $u(\xi,t)$ in the separation form of variables
\begin{eqnarray}
u(\xi,t)=\phi (\xi)e^{-i\mu t},
\end{eqnarray}
and transform Eq. (5) to the ordinary differential equation
\begin{eqnarray}
\mu \phi=- \frac{d^2\phi}{d
\xi^2}+[V_0\cos(2\xi)+g_{1d}|\phi|^{2}]\phi,
\end{eqnarray}
where $\mu$ is a constant adjusted by the normalization condition
and can be call the chemical potential. By using the balance
technique, we establish the balance condition \cite{Hai, Hai2}
\begin{eqnarray}
g_{1d}|\phi|^{2}+V_0\cos(2\xi)=\mu-1,
\end{eqnarray}
and then reduce Eq. (7) to the linear Schr\"{o}dinger equation
\begin{eqnarray}
\frac{d^2\phi}{d \xi^2}=-\phi.
\end{eqnarray}
The exact solution of Eq. (8) must obey Eqs. (9) and (10)
simultaneously. General solution of the complex equation (10) can be
written as
\begin{eqnarray}
\phi=(A+i C)\cos \xi+(B+i D)\sin \xi,
\end{eqnarray}
where $A,B,C$ and $D$ are real constants which are determined partly
by the balance condition (9). From Eq. (11) we construct the
quadratic norm
\begin{eqnarray}
|\phi|^2&=&A^2+C^2+(B^2+D^2-A^2-C^2)\sin^2 \xi\nonumber
\\&+&(AB+CD)\sin(2\xi).
\end{eqnarray}
Comparing Eq. (9) with Eq. (12) and noticing
$\cos(2\xi)=1-2\sin^2\xi$ produce the algebraical equations
\begin{eqnarray}
& & g_{1d}(A^2+C^2)=\mu-V_0-1,\ \ \ AB+CD=0,\nonumber \\
& & g_{1d}(B^2+D^2-A^2-C^2)=V_0,
\end{eqnarray}
which denote a group of indefinite equations with infinite numbers
of solutions.

The existence of multiple solutions of Eq. (13) implies that phase
of the exact solution
\begin{eqnarray}
\arctan \frac{(C\cos \xi+D\sin \xi)}{(A\cos \xi+B\sin \xi)},
\end{eqnarray}
has some arbitrariness. However, quadratic norm of the solution (7)
is determined uniquely by Eq. (9), $|\phi|^{2}
=[\mu-1-V_0\cos(2\xi)]/g_{1d}$. This means that under the exact
state the atomic density profile shapes the periodic wave-packets
which propagate with acceleration $a$. Therefore, we can achieve the
accelerated transport of BEC, through the considered exact solution.
Application of the normalization integral yields the average number
of condensed atoms per well \cite{Bronski}
\begin{eqnarray}
N'=(\pi)^{-1} \int_{0}^{\pi} |\phi(x)|^{2}dx=(\mu -1)/g_{1d},
\end{eqnarray}
which determines the chemical potential as $\mu=g_{1d}N'+1$. Given
the chemical potential, the exact atomic density reads
\begin{eqnarray}
|\phi|^{2}=R^2= N'-\frac{V_0}{g_{1d}}\cos(2\xi).
\end{eqnarray}
The result is accurately adjusted by the experimental parameters
$N',\ V_0,\ g_{1d}$, and wave vector $k_L$ and acceleration $a$
implied in $\xi$.

Clearly, Eq. (8) is a periodically driven GPE of the spatiotemporal
evolution in which the well-known Smale-horseshoe chaos exists for a
certain parameter region \cite{Chong1}. Writing the solution $\phi$
in the form of $\phi=R(\xi)\exp[i\theta(\xi)]$ and inserting it into
Eq. (8) lead to two couple equations
\begin{eqnarray}
\frac{d^2R}{d \xi^2}&=&R\Big(\frac{d\theta}{d \xi}\Big)
^2+g_{1d}R^3+[V_0\cos(2\xi)-\mu]R,
\\
\frac{d^2\theta}{d \xi^2}&+&2\theta_{\xi} \frac{R_{\xi}}{R}=0.
\end{eqnarray}
The square of the modulus $|\psi|^2=|\phi|^2=R^2$ denotes atomic
number density and the total phase reads
\begin{equation}
\Theta (x,t)=\theta(\xi)-(\mu+\alpha n\pi) t-(\alpha x t+\frac 1 3
\alpha^2t^3).
\end{equation}
The both are associated with the velocity field $v$ and  flow
density $J$, through the formulas $v=\hbar \Theta_{x}/m$ and
$J=vR^2$. Integrating Eq. (18) yields the part phase
\begin{equation}
\theta=\int \frac{J_0}{R^2}\ d\xi.
\end{equation}
So the velocity field and flow density become
\begin{eqnarray}
v(\xi,t)&=&\frac{\hbar k_L}{m}(\theta_\xi-\alpha t)=\frac{\hbar
k_L}{m}\Big[\frac{J_0}{R^2(\xi)}-\alpha t\Big],\nonumber \\
J(\xi,t)&=&\frac{\hbar k_L}{m}[J_0-\alpha R^2(\xi)t],
\end{eqnarray}
where $J_0=\theta_{\xi}(\xi_0)R^2(\xi_0)$ denotes an integration
constant determined by the flow density at the point
$\xi_0=x_0+\alpha t_0^2$ for the initial time $t_0$ and boundary
coordinate $x_0$. Substituting Eq. (16) into Eq. (21) produces the
exact flow velocity and flow density. The linear term of time in
flow velocity implies the BEC superfluid being accelerated.

\section{Regular and chaotic numerical solutions}

In general, the balance condition (9) cannot be satisfied such that
we have to solve Eq. (17) for the modulus $R$. Applying
$\theta_\xi=J_0/R^2$ of Eq. (20) to Eq. (17), we arrive at the
decoupled equation
\begin{equation}
\frac{d^2R}{d
\xi^2}=\frac{J_0^2}{R^3}+g_{1d}R^3+[V_0\cos(2\xi)-\mu]R.
\end{equation}
This is a real equation on the accelerated reference frame, which
can be reduced to the parametrically driven Duffing equation
\cite{Parthasarathy, Venkatesan} for the case $J_0=0$. The
Smale-horseshoe chaos in $J_0=0$ case has been widely investigated.
Generally, $J_0 \neq 0$, the system becomes more complicated and the
chaotic property may be kept. Under balance condition (9) we can
prove directly that Eq. (16) is an exact special solution of Eq.
(22) for a particular integral constant. When the periodic driving
is weak enough, the chaotic perturbed solution has been constructed
for the system without linear potential \cite{Chong1,Chong2}. For
most of parameters and initial data Eq. (22) is not analytically
solvable, which necessitates the numerical solutions.

At the initial and boundary point $\xi_0$, the values of the
velocity field and number density can't be determined exactly in
experiment due to the fluctuation of the atomic thermal cloud, we
can randomly choose some possible values to perform the numerical
calculations. In order to explore the analytically insolvable system
(22), we numerically solve it by using the MATHEMATICA code
\begin{eqnarray}
&&T=\pi;e[\{Rnew_{-},vnew_{-}\}]:=\{R[T],v[T]\}/.Flatten \nonumber
\\ &&[NDSolve[\{R'[\xi]==v[\xi],v'[\xi] ==g_{1d}R[\xi]^3-\mu R[\xi]\nonumber \\
&&+J_0^2/R[\xi]^3+V_0 Cos[2\xi]*R[\xi],R[0]==Rnew,\nonumber
\\
&& v[0]==vnew\},\{R,v\},\{\xi,0,T\}]];Do[pic_i=ListPlot\nonumber
\\ && [Drop[Nestlist[e,\{Random[Real,\{-0.5,0.5\}],\nonumber \\ &&Random[Real,\{-0.5,0.5\}]\},5100],100],\{i,1,10\}]
\nonumber
\end{eqnarray}
to make $10$ groups of orbits on the Poincar\'{e} section of the
equivalent phase space $(R, R_\xi)$ for the random initial
conditions $\{R(\xi_0)\in[-0.5,0.5],\ R_\xi(\xi_0)\in[-0.5,0.5]\}$
and different parameter sets. Each of the groups contains $10$
orbits, which corresponds to one of the following cases:

\begin{figure*}[htp] \center
\includegraphics[width=1.5in]{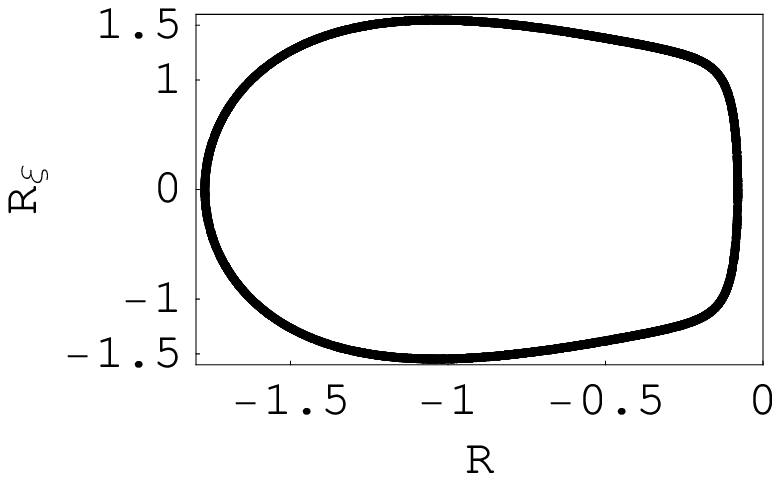}
\includegraphics[width=1.5in]{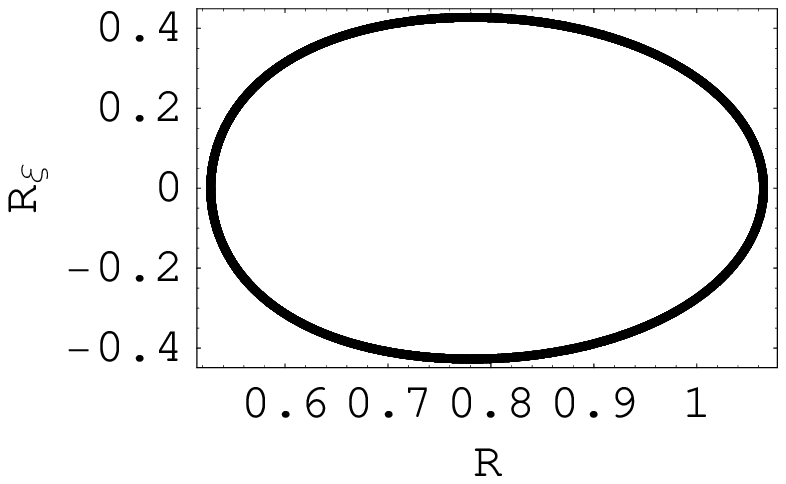}
\includegraphics[width=1.5in]{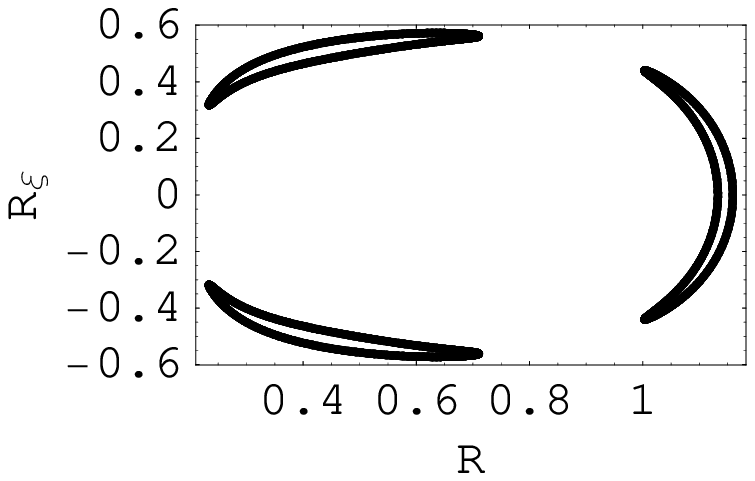}
\includegraphics[width=1.5in]{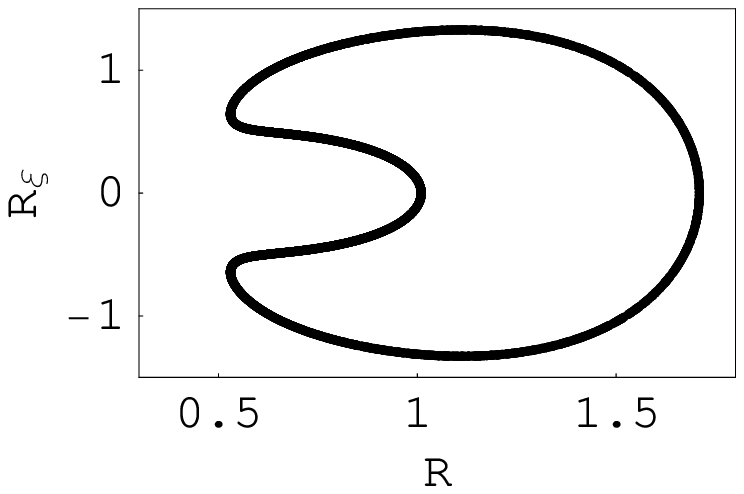}
\caption{Typical regular orbits on the Poincar\'{e} sections of the
dimensionless `phase space' $(R,\ R_\xi)$ for ten different initial
conditions and/or parameter sets. Here the phase orbits evolve in a
finite region and display the regular state distributions.}
\end{figure*}

\begin{figure*}[htp] \center
\includegraphics[width=1.5in]{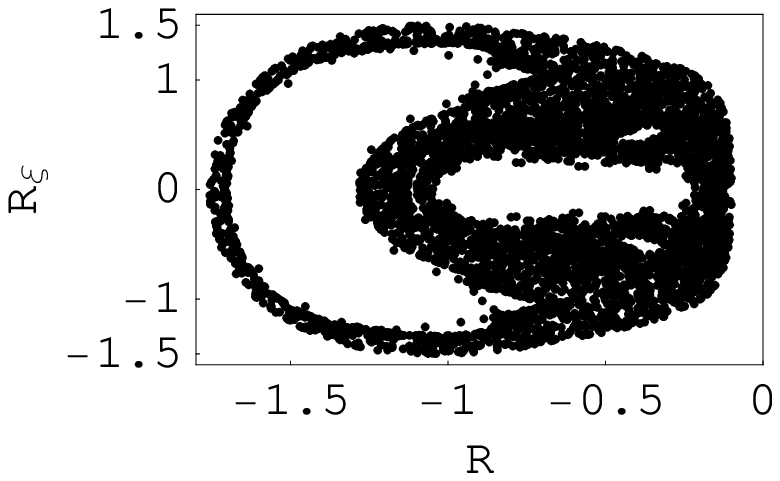}
\includegraphics[width=1.5in]{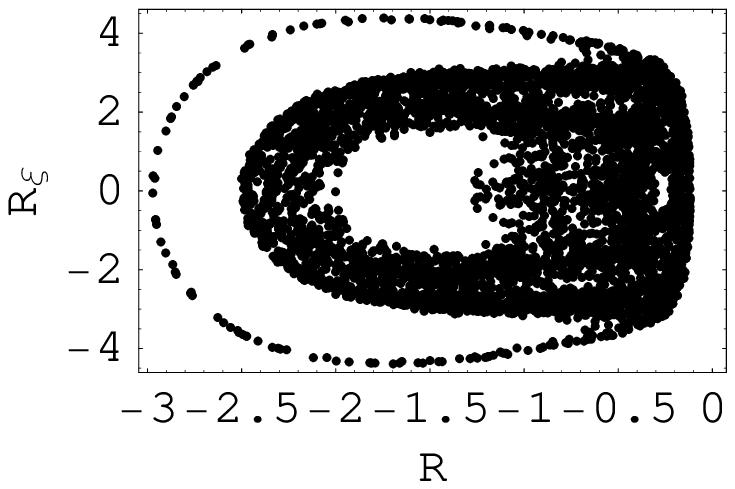}
\includegraphics[width=1.5in]{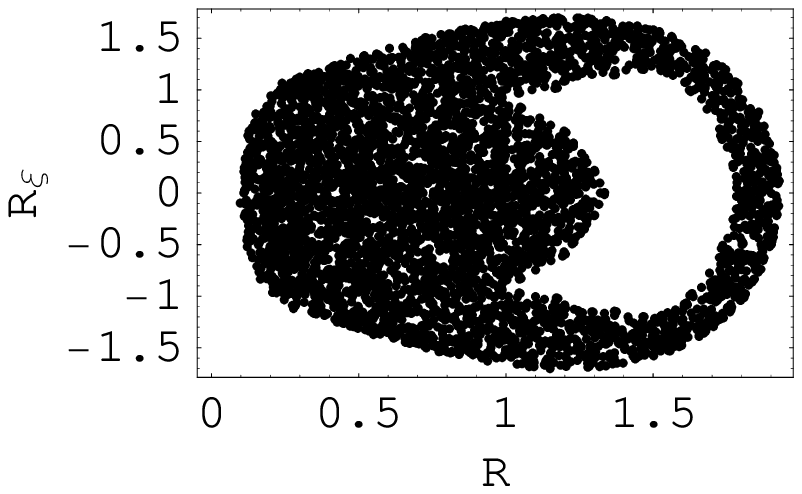}
\includegraphics[width=1.5in]{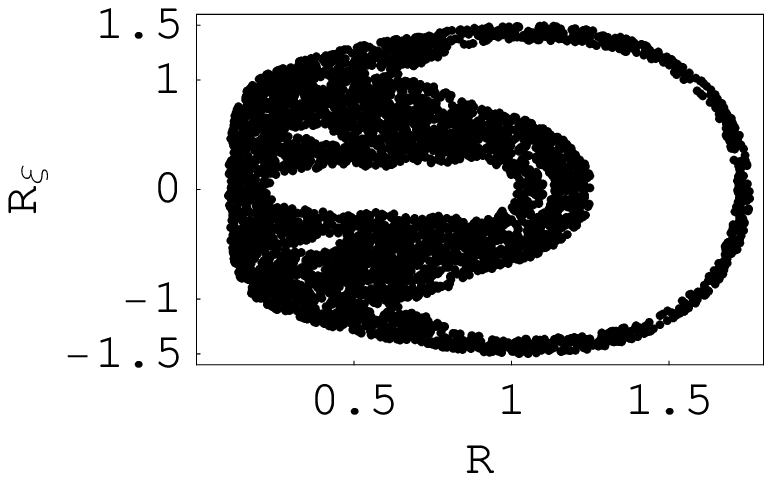}
\caption{Typical chaotic orbits on the Poincar\'{e} sections of the
dimensionless `phase space' $(R,\ R_\xi)$ for ten different initial
conditions and/or parameter sets. Here the phase orbits evolve in a
finite region and exhibit the confused state distributions.}
\end{figure*}

Case 1. For the parameter set $g_{1d}=-1,\ \mu=-0.5,\ J_0=0.01,\
V_0=0.05$ all the $10$ orbits are regular, whose $3$ typical
profiles are shown in Fig. 1.

Case 2. By increasing the strength of lattice potential to $V_0=0.2$
and keeping the other parameters as in case 1, we observe $3$
regular orbits of Fig. 1 and $7$ chaotic orbits, whose $2$ typical
profiles are shown in Fig. 2.

Case 3. After further increasing the strength of lattice potential
to $V_0=0.5$, all the orbits become chaotic.

Case 4. By increasing the flow density to $J_0=0.16$ and keeping the
other parameters as in case 3, we find that all the orbits become
regular.

Case 5. By increasing the strength to $V_0=3$ and keeping the other
parameters as in case 4, we observe $4$ regular and $6$ chaotic
orbits.

Case 6.  After further increasing the strength to $V_0=5$, all the
orbits become chaotic.

Case 7. For the parameter set $g_{1d}=-1,\ \mu=0.5,\ J_0=0.16,\
V_0=0.5$ with positive chemical potential $\mu$ all the chaotic
orbits in case 3 are transformed to regular ones.

Case 8. By increasing the strength to $V_0=2$ and keeping the other
parameters as in case 7, we observe $2$ regular and $8$ chaotic
orbits.

Case 9. After further increasing the strength to $V_0=4$ and keeping
the other parameters as in case 8, all the orbits become chaotic.

Case 10. For the different parameter sets with positive interaction
$g_{1d}$ we make many regular orbits and no chaotic orbit is found.

The above results imply that for the attractive interaction and
negative chemical potential increasing the lattice strength can
strengthen chaoticity of the system. Conversely, increasing the flow
density can weaken the chaoticity. On the other hand, after changing
the chemical potential from negative to positive, the chaoticity is
weakened effectively. Finally, for the positive interaction $g_{1d}$
we have not found the chaotic orbits. The $4$ typical regular orbits
and $4$ typical chaotic orbits among the $10\times 10=100$ orbits
are shown in Fig. 1 and Fig. 2 respectively. They are associated
with different initial conditions and/or parameter sets. As the
examples of densities of atomic number we plot their spatiotemporal
evolutions for the parameters of Case 2 and two fixed initial
conditions as in Fig. 3. Figure 3(a) corresponds to the first closed
orbit of Fig. 1 and describes a quasiperiodic evolution thereby.
Figure 3(b) is associated with the first chaotic orbit of Fig. 2,
which possesses obvious aperiodicity. From Eq. (21) the
corresponding flow densities can be easily illustrated. They will
increase linearly in time that leads the condensed atoms to tunnel
out of the optical traps.
\begin{figure}[htp] \center
\includegraphics[width=2.0in]{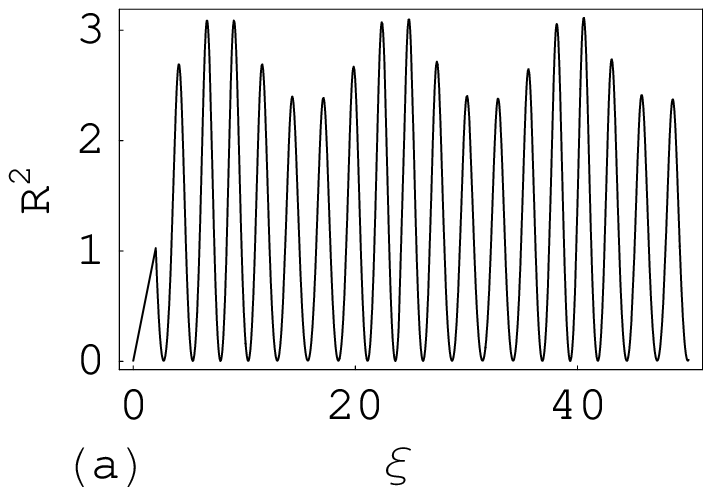}
\includegraphics[width=2.1in]{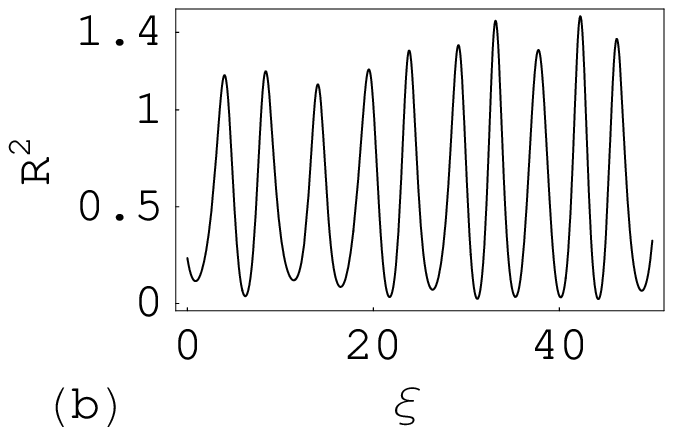}
\caption{Spatiotemporal evolutions of the densities of atomic number
for the parameters of Case 2 and the initial conditions (a)
$R^2(0)=-0.07793183579,\ R_\xi^2(0)=-0.199975080313$, (b)
$R^2(0)=-0.48451144892,\ R_\xi^2(0)=0.31792019031$.}
\end{figure}
These results reveal the relations between the system parameters and
the solution behaviors, and suggest a method for controlling the
regular and chaotic states.

\section{Conclusions}

In summary, we have investigated a BEC interacting with an
accelerated WS potential. Using the mean-field method and the
macroscopic one-body wave function, we seek the exact analytical
solution and the regular and chaotic numerical solutions of the
system. It is demonstrated that after the linear potential being
removed by a function transformation, the governing equation becomes
a periodically driven GPE on the accelerated reference frame. With
the help of the balance condition, we establish the first exact
analytical solution of the WS system, which is accurately controlled
by the experimental parameters. Further writing the solution of GPE
in the exponential form, we obtain the equation of modulus, which
contains the parametrically driven Duffing equation. The well-known
Smale-horseshoe chaos and quasiperiodic orbits on the Poincar\'{e}
sections of the dimensionless `phase space' are shown numerically
for different initial conditions and parameter regions of different
chaoticity. The accelerated atomic flow densities are demonstrated
for both the regular and chaotic states.

It is well known that chaos could emerge in the processions of BEC
collapsing and may play a destructive role for the BEC system.
Therefore, predicting and controlling chaos are quite important for
the creation and application of BEC. Our analytical and numerical
results have supplied a method for controlling the regular and
chaotic states, through the application of accelerated optical
potential and the adjustments of system parameters.

\begin{acknowledgments}
This work was supported by the National Natural Science Foundation
of China under Grant Nos. 10575034 and 10875039.
\end{acknowledgments}

\end{document}